\documentclass[linenumbers,twocolumn,twocolappendix]{aastex631}
\nolinenumbers

\usepackage[ruled]{algorithm2e}
\usepackage{enumitem}
\usepackage{amsmath}
\usepackage{booktabs}
\usepackage{listings}
\usepackage[newfloat,frozencache,cachedir=minted-output]{minted}
\setminted[C]{fontsize=\footnotesize}
\BeforeBeginEnvironment{minted}{\medskip}
\AfterEndEnvironment{minted}{\medskip}

\begin{document}
\watermark{preprint}
\title{Gamma-ray burst detection with Poisson-FOCuS and other trigger algorithms}

\correspondingauthor{Giuseppe Dilillo}
\email{peppedilillo@gmail.com, giuseppe.dilillo@inaf.it}
\affiliation{INAF - Istituto di Astrofisica e Planetologia Spaziali, Roma, Italy}
\affiliation{Dipartimento di Scienze Matematiche, Informatiche e Fisiche, Università di Udine, Udine, Italy}

\author{Giuseppe Dilillo}
\affiliation{INAF - Istituto di Astrofisica e Planetologia Spaziali, Roma, Italy}
\affiliation{Dipartimento di Scienze Matematiche, Informatiche e Fisiche, Università di Udine, Udine, Italy}

\author{Kes Ward}
\affiliation{STOR-i Doctoral Training Centre, Lancaster University, Lancaster, UK}

\author{Idris A. Eckley}
\affiliation{Department of Mathematics and Statistics, Lancaster University, Lancaster, UK}

\author{Paul Fearnhead}
\affiliation{Department of Mathematics and Statistics, Lancaster University, Lancaster, UK}

\author{Riccardo Crupi}
\affiliation{Dipartimento di Scienze Matematiche, Informatiche e Fisiche, Università di Udine, Udine, Italy}

\author{Yuri Evangelista}
\affiliation{INAF - Istituto di Astrofisica e Planetologia Spaziali, Roma, Italy}
\affiliation{INFN - sezione Roma Tor Vergata, Rome, Italy}

\author{Andrea Vacchi}
\affiliation{Dipartimento di Scienze Matematiche, Informatiche e Fisiche, Università di Udine, Udine, Italy}
\affiliation{INFN - Sezione di Trieste}

\author{Fabrizio Fiore}
\affiliation{INAF - Osservatorio Astronomico di Trieste, Italy}
\affiliation{IFPU: Institute for the Fundamental Physics of the Universe, Trieste, Italy}

\begin{abstract}
We describe how a novel online changepoint detection algorithm, called Poisson-FOCuS, can be used to optimally detect gamma-ray bursts within the computational constraints imposed by miniaturized satellites such as the upcoming HERMES-Pathfinder constellation. Poisson-FOCuS enables testing for gamma-ray burst onset at all intervals in a count time series, across all timescales and offsets, in real-time and at a fraction of the computational cost of conventional strategies.
We validate an implementation with automatic background assessment through exponential smoothing, using archival data from Fermi-GBM. Through simulations of lightcurves modeled after real short and long gamma-ray bursts, we demonstrate that the same implementation has higher detection power than algorithms designed to emulate the logic of Fermi-GBM and Compton-BATSE, reaching the performances of a brute-force benchmark with oracle information on the true background rate, when not hindered by automatic background assessment.
Finally, using simulated data with different lengths and means, we show that Poisson-FOCuS can analyze data twice as fast as a similarly implemented benchmark emulator for the historic Fermi-GBM on-board trigger algorithms.
\end{abstract}


\section{Introduction}\label{sec1}

Gamma-ray bursts (GRBs) manifest as sudden, transient increases in the count rates of high-energy detectors, see e.g. the review of \cite{meszaros2018gamma}, \cite{berger2014short} and \cite{nakar2020electromagnetic}. These events appear as unexpected, their activity not explainable in terms of background activity or other known sources. 
At a fundamental level, algorithms for GRB detection stayed the same through different generations of spacecrafts and experiments: as they reach the detector, high-energy photons are counted in different energy bands; an estimate of the background count rate is assessed from past observations; the observed photon counts are compared against the number of photons expected from the background, at regular intervals and over multiple, predefined timescales. 
The latter step is justified by a physical reason: GRBs have been observed with much different duration and lightcurves. Since the degree of similarity between the duration of a GRB and the timescale that is tested affects the power of the test, by testing multiple timescales the chance of a burst being missed due to a mismatch is reduced.
The process is iterated until eventually a large enough excess in the observed count is found, relative to the counts expected from the background. The specificity of the search is regulated through a threshold parameter, usually expressed in units of standard deviations.
When the threshold is reached, a quality assurance step may take place to prevent false detection, such as accepting a trigger only if it is observed across multiple detectors. For space-borne observatories it is critical that algorithms for GRB detection run online i.e., at the same time in which data are collected. A timely alert may in fact serve to trigger the acquisition of a recording apparatus or to initiate follow-up observations from the ground.

Over time, \emph{trigger algorithms} have grown to support an increasing number of criteria and parameters to achieve greater sensitivity to different bursts. While Compton-BATSE tested for three timescales \citep{1999paciesas, 1999kommers}, more than 120 and 800 different trigger criteria can be specified for Fermi-GBM \citep{meegan2009fermi} and Swift-BAT \citep{mclean2004setting} respectively.
The Compton-BATSE's algorithm performs a subset of the test that are performed by Fermi-GBM, hence their performances have been compared in literature. 
\cite{paciesas2012fermi} reports that a minimum of $83$ GRBs out of the $491$ bursts which triggered the Fermi-GBM algorithm during the first two years of operation, were detected over timescales not tested by the Compton-BATSE's algorithm. Most of these events were identified as long GRBs and were detected over timescales greater than $1024$~ms. A single GRB was detected over an energy range outside the standard BATSE energy range ($50$--$300$ keV), at timescales already covered by the BATSE algorithm. Analyses of GRBs discovered by Fermi-GBM in the third and fourth years of operation yielded similar results \citep{von2014second}. This suggests that the range of timescales tested for a change is the most important feature of GRB detection algorithms.
Growth in algorithm complexity is not without consequences. The simultaneous operation of many trigger criteria can require most of the available CPU time \citep{fenimore2003trigger} and trigger algorithms have been simplified during scientific operations to reduce the computational burden on on-board computers \citep{paciesas2012fermi}.
Alternative approaches exist. For example, \cite{scargle2013studies} discuss the application of a multiple changepoint model called Bayesian Blocks to the problem of online GRB detection. Bayesian Blocks overcomes many of the limitations of conventional algorithms, such as the need to specify testing timescales. However, due to its quadratic computational cost, the technique faces even more severe performance challenges than established solutions.
This picture is unfavorable to nanosatellite missions due to various constraints specific to these spacecrafts, such as miniaturized computers, limited electrical power systems, and the absence of active temperature control, which result in a limitation of the available computational resources.

We investigate the application of a recently developed changepoint detection technique to make a more accurate, faster and simpler algorithm for GRB detection. 
FOCuS was introduced in \cite{romano2023fast}. 
The first version of the algorithm was designed for detecting anomalies in normally distributed data. In \cite{ward2022}, FOCuS was extended to Poisson-distributed data (Poisson-FOCuS). A strategy to reduce the computational costs from linearithmic to linear was devised in \cite{ward2023constant}. In essence, Poisson-FOCuS enables testing all the intervals in a count time series for transient onset, that is testing over all timescales and offsets, at the same time in which data are collected and at a fraction of the computational cost of conventional strategies.
The technique builds upon methods with a long history of applications: the mathematical backbone of FOCuS is the CUSUM test, a classic technique for anomaly detection \citep{page1954continuous, lucas1985counted}; and the algorithm's mechanics can be traced back to fundamental algorithms in computational geometry, such as Graham's scan \citep{graham1972efficient}.

Poisson-FOCuS was specifically developed to detect GRBs, in an effort to identify an algorithm for the satellites of the oncoming HERMES-Pathfinder and SpIRIT nanosatellite constellation \citep{fiore2020hermes}.
HERMES-Pathfinder is an in orbit demonstration using six 3U CubeSat, with the goal of demonstrating that GRBs can be detected and localized by miniaturized instruments hosted by nano-satellites (\url{www.hermes-sp.eu}). 
HERMES Pathfinder is funded by the Italian Space Agency and the European Commission through a H2020 grant. SpIRIT is a 6U CubeSat funded by the Australian Space Agency, managed by University of Melbourne and hosting one HERMES-Pathfinder payload unit \citep{thomas2023localisation}. 
SpIRIT is due to launch at the end of November 2023. HERMES Pathfinder should reach orbit during 2024.

This paper is divided into two parts. In the first section, we introduce a framework for evaluating, comparing, and visualizing the operations of algorithms for GRB detection. We provide an account of the Poisson-FOCuS algorithm and its implementation, based on a design proposed for application on the satellites of the HERMES-Pathfinder constellation.
In the second part of our study, we assess the performance of different algorithms using both real and simulated data. Using simulated data, we evaluate the detection power of an implementation of Poisson-FOCuS with automatic background assessment through simple exponential smoothing (FOCuS-AES). Results are compared to those obtained with algorithms that mimic the logic of Fermi-GBM and Compton-BATSE, as well as a brute-force benchmark with oracle information on the background--that is,  access to the true background count rate used in the simulations. 
In these tests, FOCuS-AES achieved detection performances similar to the benchmark when not limited by the automatic background assessment. On the other hand, FOCuS and the benchmark also resulted in higher false-positive rates than less sensitive algorithms.
In a second experiment, we test the computational performances of Poisson-FOCuS against that of a similarly implemented benchmark emulator for the Fermi-GBM trigger logic. Poisson-FOCuS completed these tests in approximately half the time required by the benchmark.
Finally, we evaluate FOCuS-AES over three weeks of real data from Fermi-GBM, selected at different times during the mission. All GRBs previously detected by Fermi-GBM were correctly identified, as well as a number of transients with counterparts in the untriggered GBM Short GRB candidates catalog \citep{fermiuntriggered}. However, FOCuS-AES also detected a number of transients with no counterparts in GBM catalogs, likely originating from the Sun, during periods of high solar activity.
All the data and code used in this research are publicly available.

\section{Algorithms for GRB detection}
\label{sec:trigal}
Consider the problem of detecting a GRB in the following form. Over time, high-energy photons reach a count detector on-board a spacecraft.
Photons are counted within discrete time steps of duration $\tau$ (the bin length), resulting in a collection of $t$ values after $t$ time steps, or after a time $t\tau$ from the start of the observations.
We denote by $x_i$ the $i$-th value of the count time series, and the count time series comprising the first $t$ values by $X_t = \{x_1,..,x_t\}$.
Tests for a GRB involve comparing the observed and expected number of photons over different intervals of time. Each interval in the time series can be identified by two indices, $\{i, h\}$, representing the interval's ending index and its length, respectively. The total number of counts in a given interval is noted with a lowercase letter, $x_{i,h} = \sum_{j = i - h + 1}^{i} x_j$.
At the $t$-th time step, there are a total of $t(t + 1)/2$ unique intervals, see for example the bottom panel of Fig.~\ref{fig:checkers_gbm}. At the next time step, the number of unique intervals increases by $t + 1$. All these new intervals span the most recently acquired count $x_{t + 1}$. In our notation, they are represented by $\{t+1, h\}$ with $0 < h \leq t + 1$.

During most of the observation time, photons reaching the detector are stochastically and predictably emitted by a large number of unknown background sources.
We assume that a reasonable estimate of the background count rate $b_i$ is available each time a newly observed count value $x_i$ is recorded. The background estimates are represented in a parallel time series, $B_t = \{b_1,..,b_t\}$.
The same notation mentioned above is adopted for the background, so that the total number of photons expected from all background sources over the interval $\{i,h\}$ is $b_{i,h} = \sum_{j = i - h + 1}^{i} b_j$.

Suddenly, a source in the detector's field of view shines brightly, leading to an unexpected increase in the observed counts. A significance score $S_{i,h}$ is associated with each interval. The significance measures the `extraordinariness' of the number of photons collected relatively to the number of photons expected from background, hence $S_{i,h} = S(x_{i,h}, b_{i,h})$. Since we are interested in bright transients, we consider only intervals for which $x_{i,h} > b_{i,h}$. However, it is straightforward to adapt our discussion to scenarios in which intensity deficits are pursued, e.g., when $x_{i,h} < b_{i,h}$--think, for example, of occultation phenomena.

We investigate automatic techniques to detect intervals of the count time series with excess significance greater than a predefined threshold expressed in terms of standard deviations $S_{i,h} > T$. We call these techniques trigger algorithms since a positive detection may serve to trigger the operation of secondary apparatus, such as a high-resolution recording system or the observation of a small-field telescope.
We restrict ourselves to strategies which can be run online, i.e. sequentially and over time series of arbitrary length.  In practice, this requires that after a number of iterations $t$, no more than the first $t$ elements of the count and background time series shall be accessible to the method.
We assume observations to be collected for a single energy range and detector. This is reasonable when the number of detector-range combinations is small. Indeed, to widen the search to multiple energy bands and detectors, different algorithm instances can be stacked with costs only scaling linearly. 
Finally, we assume that algorithms have access to an estimate of the count rate expected from all background sources at each iteration.  We do not require the background mean rate to be constant in time, although in practice background rates are most often assumed to change slowly relative to the transients' duration. We will return on the subject of estimating count rate from background sources in Appendix \ref{apx:background}.

\subsection{Exhaustive search}
\label{sec:exh}

It is trivial to design an exhaustive search algorithm for solving the GRB detection problem: at the $t$-th iteration step all the intervals $\{t, h\}$ with duration $h$ such that $0< h \leq t$ have their significance score $S_{t,h}$ computed and tested against the threshold $T$, see Alg. \ref{algo:exhaustive}. Since the total number of intervals in an observation series $X_t$ is $t(t+1)/2$, the computational cost of an exhaustive search algorithm grows as the square of the number of observations $t$, $O(t^2)$. For this reason, exhaustive search algorithms are of no practical interest for real-world applications.

Many of the significance tests performed by exhaustive search algorithms (and by extension, conventional algorithms) can be avoided. This point is best illustrated through a simple numerical example. An exhaustive search algorithm with a standard deviation threshold $T > 0$ is executed over the count observations time series $X_2 = \{90, x_2\}$ and the background count rates $B_2 = \{100.0, 100.0\}$. This implies that after the second iteration, three significance computations $S_{1,1}$, $S_{2,1}$ and $S_{2,2}$ have been performed. 
If $x_2>100$ then $S_{2,2} < S_{2,1}$, hence either $S_{2,1} > T$ or no trigger is possible at all, making the actual computation of $S_{2,2}$ unnecessary regardless of the $x_2$ value.
Indeed, exhaustive search algorithms make no use of the information gained during the algorithm operations. This information can be used to avoid significance computations for intervals which can not possibly result in a trigger.
Regardless of their computational performances, an exhaustive search algorithm is a useful benchmark against which the detection performances of other techniques can be evaluated. The detection power of an exhaustive search algorithm for GRB detection is in fact maximum, meaning that it would always be able to meet the trigger condition over the earliest intervals whose significance exceeds the algorithm threshold. An exhaustive search running over data for which the background count rate is known is biased only by the fundamental binning of the data.

\begin{algorithm}
\caption{A sketch of an exhaustive search algorithm for GRB detection.}
\label{algo:exhaustive}
\While{\textbf{not} \upshape $\texttt{triggered}$}{
	\For{$h \leq t + 1$}{
		compute $S = S_{t,h}$\;
		\If{$S > T$}{
			\texttt{triggered} $\leftarrow$ \textbf{True};
			}
	}
    \texttt{t} $\leftarrow$ \texttt{t + 1};
}
\end{algorithm}

\subsection{The conventional approach to GRB detection}
\label{sec:partrig}

\begin{figure*}
    \centering
	\includegraphics[width=0.78\linewidth]{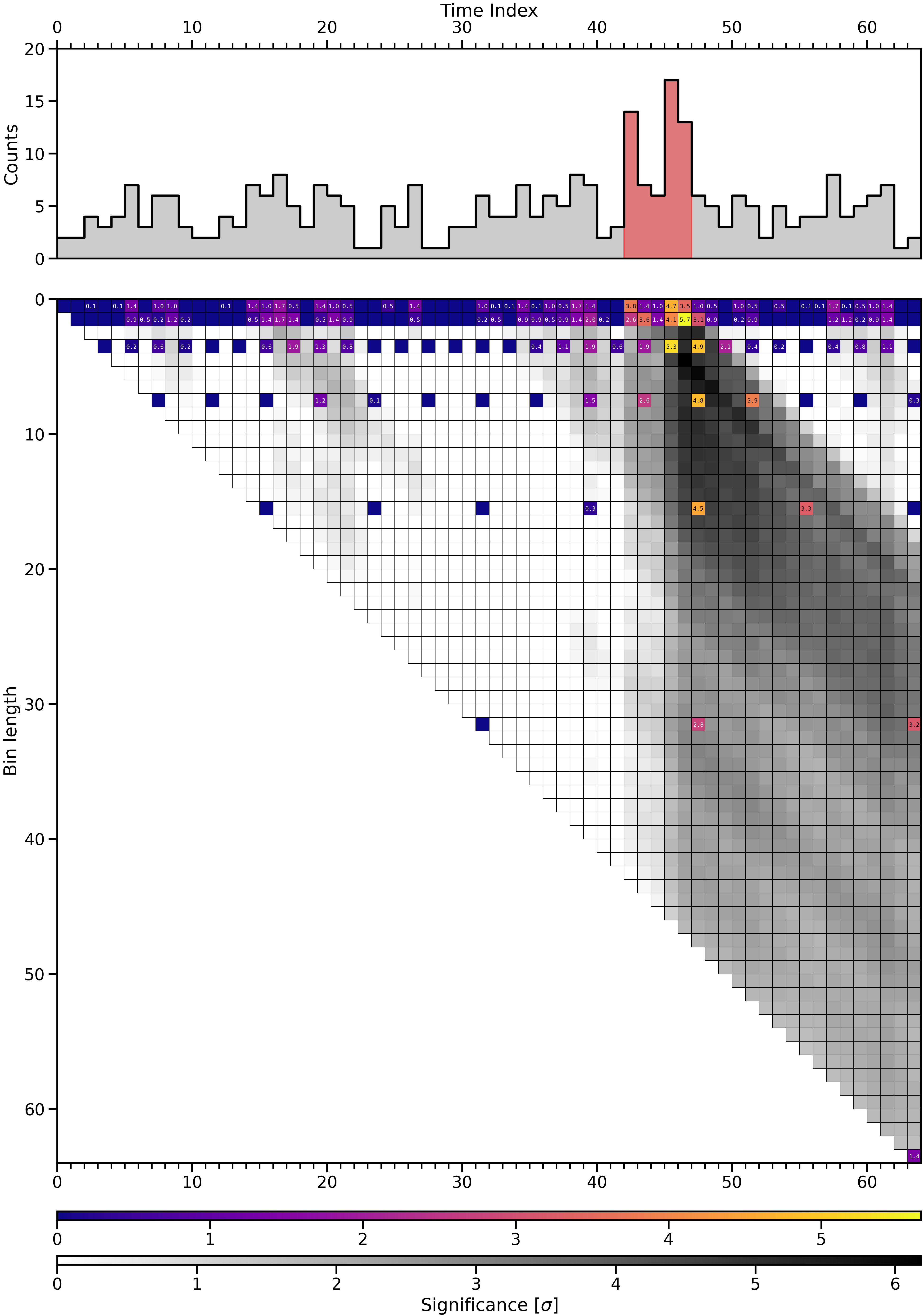}
	\caption{
    Operations of a conventional algorithm for detecting GRBs, testing logarithmically equispaced timescales at offsets equal to half the bin-length. 
    A time series of counts is represented in the top panel. All the counts but those in the interval $\{i=47, h=5\}$ (or equivalently, $41 < t \leq 47$) are sampled from a Poisson distribution with mean rate $\lambda = 4.5$ (background). 
    Counts in the interval $\{i=47, h=5\}$ are sampled from a Poisson distribution with mean rate $\lambda = 9.0$ (transient; red, shaded). In the bottom panel, every interval of the time series $\{i,h\}$ is represented with a tile. 
    Longer intervals are found at the bottom, early intervals are at the left. The shade of a tile represents the significance of the excess of count relative to the background mean, in units of standard deviations. Intervals tested by the algorithm are represented with colored tiles, while shades of grey are used for the remaining intervals. This figure is meant to be displayed in color.
    }
    \label{fig:checkers_gbm}
\end{figure*}

The logic of a conventional algorithm is sketched in Alg. \ref{algo:trad}. Most of the algorithms (including those discussed in the introduction) assisting instruments monitor on-board real spacecrafts belong to this category.
These algorithms perform a grid search over time. This means that tests of a predefined timescale are repeated at regular intervals. The search stops when an interval significant enough is eventually found. If an interval with significance $S_{i,h} > T$  exists, there is no guarantee that a conventional algorithm operating at threshold $T$ will result in a trigger. For this to happen, either the burst timing should match the grid search or multiple, significant enough intervals must exist. 
For a bright source, this implies that the duration of the event must be similar to the interval being tested. Alternatively, the source should be sufficiently bright to be detectable over an interval shorter or longer than its actual duration.
In other words, detections from conventional algorithms are biased towards events whose timings and durations match those of the grid search. The per-iteration computational costs of conventional algorithms is limited by a constant. This is trivially true, since the number of significance tests performed can not exceed the number of tested criteria.
The operations of a conventional algorithm checking logarithmically equispaced timescales at offsets equal to half acquisition length are represented in Fig.~\ref{fig:checkers_gbm}.

\begin{algorithm}
\caption{Sketch of a conventional algorithm for GRB detection. The maximum significance in excess counts is computed over a grid of observation intervals with timescales $h \in H$ and offsets $g \in G$.}
\label{algo:trad}
\While{\textbf{not} \upshape $\texttt{triggered}$}{
	\For{$h, g \in (H, G)$}{
	\If{$t\%h = g$}{
		compute $S = S_{t,h}$\;
		\If{$S > T$}{
			\texttt{triggered} $\leftarrow$ \textbf{True};
			}
		}
	}
    \texttt{t} $\leftarrow$ \texttt{t + 1};
}
\end{algorithm}

\subsection{Computing significances}

\begin{figure}
  \centering
    \includegraphics[width=\linewidth]{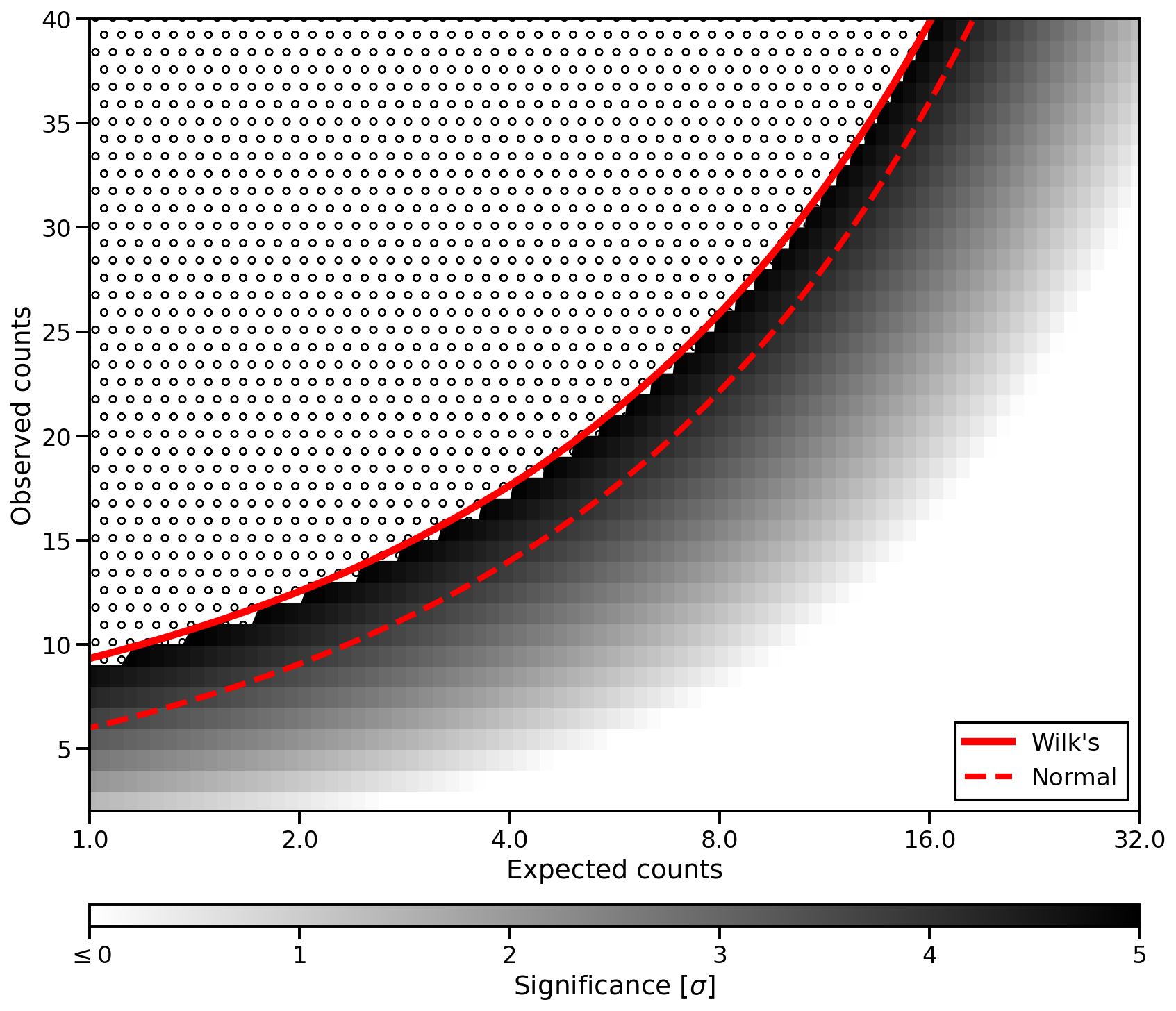}
    \caption{Threshold values according to Wilk's theorem approximate well the true value, even when the expected background count is small. The x-axis and y-axis represent different Poisson processes mean rates and observed count values, respectively. Shades of grey are used to represent the significance of excess counts, computed according to exact Poisson tail distribution probability values and expressed in units of standard deviations. The hatched region's significance exceeds $5\sigma$. A solid line is used to represent the significance threshold according to Eq. \ref{eq:wilk_sign}, while a dashed line represents the standard score threshold (normal approximation). 
}
    \label{fig:thresholds}
\end{figure}

\label{sec:statsig}
Neglecting uncertainty in background estimation, the probability of observing more than $x_{i,h}$ photons when $b_{i,h}$ photons are expected is given by the Poisson cumulative distribution function. It is common to convert this probability value to a sigma-level by calculating how many standard deviations from the mean a normal distribution would have to achieve the same tail probability. As this can be expensive to compute, the use of approximate expressions is commonplace in practice. For example, the algorithms of Swift-BAT use a recipe based on the standard score modified to support ``a commandable control variable to ensure that there is minimum variance when observed counts are small" \citep{fenimore2003trigger}. Using standard score for computing significance when few counts are expected from background sources may lead to a higher rate of false detections which is likely the reason why Swift-BAT uses dedicated minimum variance parameters. 

An alternative approach for computing the significance of an interval involves performing statistical tests of the hypotheses:
\begin{enumerate}
    \item $\textbf{H}_0$, $x_{i,h} \sim P(b_{i,h})$, null hypothesis.
    \item $\textbf{H}_1$, $x_{i,h} \sim P(\mu b_{i,h})$ with $\mu > 1$.
\end{enumerate}
where $P$ denotes the Poisson distribution and $\mu$ is the intensity parameter. Since we are presently interested in bright anomalies such as GRBs, only values of intensity $\mu$ greater than $1$ are tested. The log-likelihood ratio for this test is:
\begin{equation*}
    \lambda_{\text{LR}} = 2\Big[x_{i,h} \log\Big(\frac{x_{i,h}}{b_{i,h}}\Big) - (x_{i,h} - b_{i,h})\Big]
\end{equation*}
According to Wilk's theorem, the log-likelihood ratio $\lambda_{\text{LR}}$ is asymptotically distributed as a chi-squared distribution $\chi_d^2$, with $d$ equal to the difference in degrees of freedom between the null and alternative hypotheses. In our problem, the only degree of freedom comes from the intensity parameter $\mu$, so $d=1$. Using the relationship between the chi-squared and normal distribution and the definition of $S_{i,h}$, we obtain the significance expression in units of standard deviations:
\begin{equation}
    \label{eq:wilk_sign}
    S_{i,h} = \sqrt{2\Big[x_{i,h} \log\Big(\frac{x_{i,h}}{b_{i,h}}\Big) - (x_{i,h} - b_{i,h})\Big]}
\end{equation}
Note that in practical application, the most efficient way to assess the trigger condition $S_{i,h} > T$ is to rescale the threshold according to $T^{\prime} \leftarrow T^{2} / 2$. This avoids the need for the square root computation. 
In the literature, some authors refer to this test as the ``deviance goodness-of-fit test" \citep{hilbe2014modeling, feigelson2022}. In Fig.~\ref{fig:thresholds}, the significance estimates obtained through Eq. \ref{eq:wilk_sign} are compared to the exact values computed according to the Poisson and the normal cumulative distributions, as well as the standard score. 
While Wilk's theorem holds for large sample sizes, it is apparent that the approximation remains accurate even when testing small counts. Unlike the results obtained with the normal approximation, Wilk's threshold upper-bounds the exact solutions, providing a more conservative estimate with a lower false positive rate.

When choosing the threshold parameter one trades off potential false detections for a loss of statistical power. The best choice depends on the specific application and must take into account the cost of a false detection. For example, a false positive could result in extra telemetry or memory for storing high-resolution data.
A trade-off must be considered also when choosing the bin length, $\tau$, since power is lost to events with duration shorter than the bin length, but shortening the bin length increases the frequency of false detections. Experimentation with historic and synthetic data is crucial in determining the best combination of these parameters \citep{mclean2004setting}.

\subsection{Poisson-FOCuS}
\label{sec:focuspois}

FOCuS (Functional Online CUSUM) is a changepoint and anomaly detection algorithm that builds upon the classic CUSUM technique \citep{page1954continuous, lucas1985counted}, which itself can be understood as a repeated, one-sided Wald's sequential probability ratio tests \citep{wald1945sequential, belanger2013detecting}. It enables the detection of anomalies in time series with the sensitivity of an exhaustive search algorithm but limited computational costs. A first version of the algorithm was presented in 2021 \citep{romano2023fast}. The algorithm has later been adapted to Poisson distributed count data (Poisson-FOCuS), specifically for detecting astrophysical transients such as GRBs and targeting applications on-board the spacecrafts of the HERMES Pathfinder constellation \citep{ward2022}. The method was then shown to be applicable to other distributions of the exponential family, and a strategy called ``adaptive maxima check" was devised to reduce computational costs from linearithmic to linear \citep{ward2023constant}. The resulting technique has both maximum detection power and optimal computational cost. In this sense, we say that the algorithm is `optimal'.

\begin{figure*}
    \centering
	\includegraphics[width=0.78\linewidth]{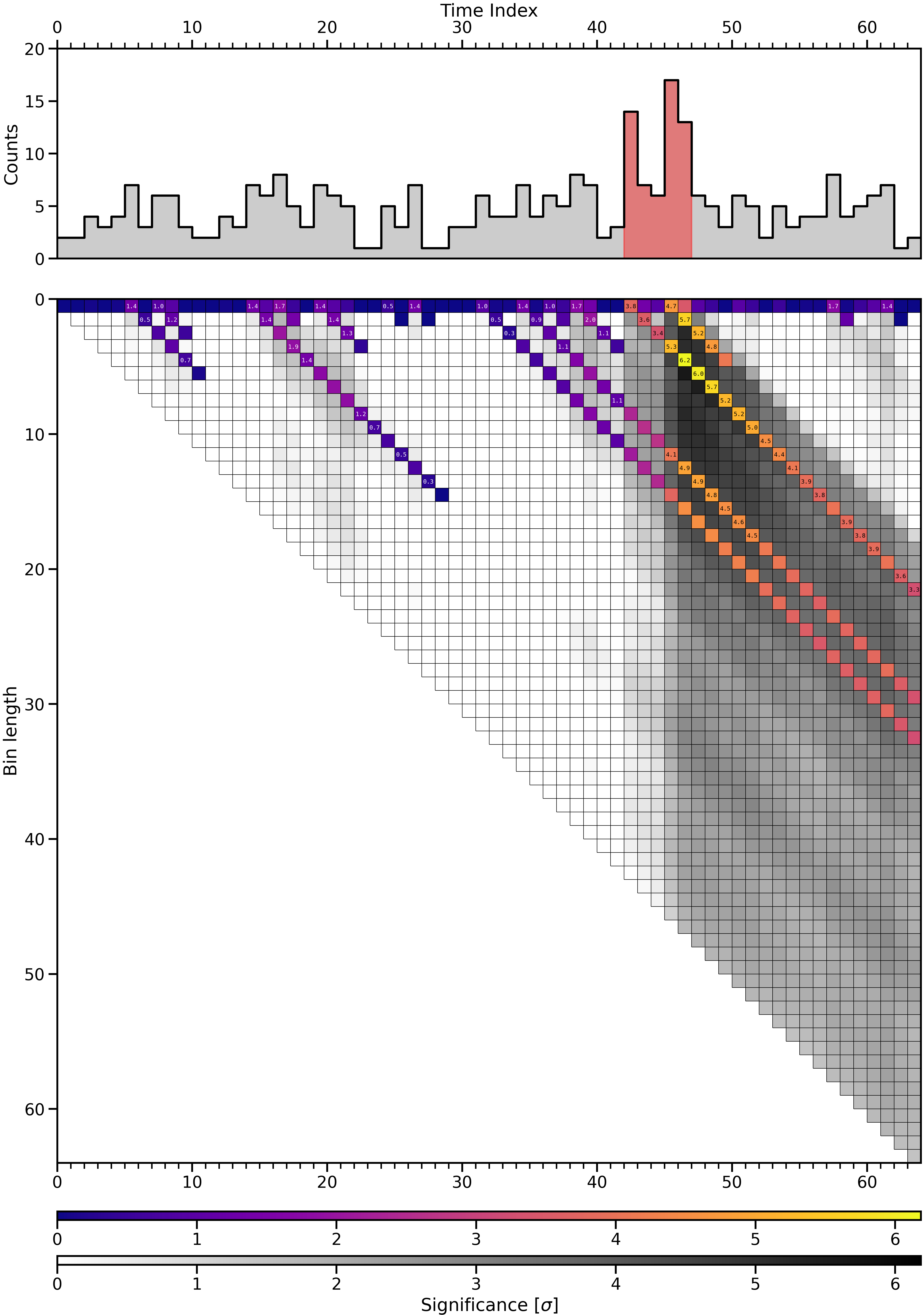}
	\caption{
    A schematic representation of the operations of Poisson-FOCuS over a simple data input, see also the caption to Fig.~\ref{fig:checkers_gbm}. A coloured tile in a column with index $i$ is associated with a curve in the FOCuS's curve stack at the $i$-th algorithm's iteration. Numeric values are used to display significance scores, only for those curves whose significance is actually computed, using adaptive maxima check with threshold $5.0 \sigma$. This figure is meant to be displayed in color.
    }
    \label{fig:checkers_focus}
\end{figure*}

Poisson-FOCuS tests for evidence of a GRB over intervals whose length is dynamically assessed by the algorithm itself, based on evidence gathered in the past. This approach contrasts with conventional strategies, which test intervals with fixed lengths, see Figure \ref{fig:checkers_focus} and compare with Figure \ref{fig:checkers_gbm}.
Poisson-FOCuS maintains a hierarchical collection called the curve stack, whose elements are called curves. Each curve corresponds to an interval $\{i, h\}$ in the time series. Curves are comparable; one curve is considered greater than another if its ``mean" $x_{i,h} / b_{i,h}$ exceeds that of the other. If a curve is smaller than an older one, it can be \emph{pruned} from the collection, as a trigger from this curve would inevitably result in a trigger from another already in the curve stack. This situation is referred to as one curve ``dominating" another.
During each iteration, the algorithm must check the significance associated with each curve. Through adaptive maxima check, an upper-bound significance is associated to each curve, and the actual significance value is computed only if the bound exceeds the threshold. When one curve significance exceeds the threshold, the algorithm returns and outputs the corresponding interval's start time (the changepoint) $i - h$, the trigger time $i$, and significance $S_{i,h}$.

The mathematics motivating Poisson-FOCuS has been comprehensively described in \cite{ward2022}. Presently, we will concentrate on the algorithmic aspects of Poisson-FOCuS and on how a robust and efficient implementation can be achieved. The code we provide is implemented in C89. This particular implementation was developed as a baseline for potential deployment on the satellites of the HERMES-Pathfinder constellation. The logic we designed implements all the optimization techniques we are aware of and does not assume the background to be constant. A minimal, functional Python implementation is presented in Appendix \ref{apx:minimalfocus}. 

The Poisson-FOCuS algorithm can be broken down into four distinct components: a constructor, an interface, the updater and the maximizer. It relies on a stack data structure, for which we do assume a standard application programming interface exists providing access to methods for iterating, \emph{pushing} and \emph{popping} from the top of the stack, and for \emph{peeking} the stack's top-most element. 

\paragraph{Curves, curve stack, accumulator}
 The curve stack is a stack data structure containing records elements called curves. 
Each curve is uniquely associated with an interval of the time series. A curve contains at least three members or fields. The count member is a non-negative integer representing the total counts observed since the curve was first added to the stack, $x_{i,h}$. The background member is a positive, real number representing the total counts expected from the background over the same interval, $b_{i,h}$. The time member is a positive integer $h$ serving as a step-index `clock'. It represents the number of iterations completed since the curve was first added to the stack. 
One extra member can be used to store an upper-bound to the curve's significance when implementing adaptive maxima check. 

Instead of updating each member of every curve in the curve stack at each iteration, Poisson-FOCuS can utilize an auxiliary curve called the accumulator. At the $i$-th algorithm iteration, the members of the accumulator are incremented by $x_i$, $b_i$ and $1$. When a curve is added to the stack it is added as a timestamp copy of the accumulator. This allows, for instance, the calculation of the counts observed since the curve's inception, $x_{i,h}$, as the difference between the accumulator and the curve's count member.

Curves are comparable; a curve $q$ dominates (is greater than) a curve $q^\prime$ if the mean of $q$ exceeds the mean of $q^\prime$:
\begin{equation}
\label{eq:dominates}
    \frac{x_{i,h}}{b_{i,h}} > \frac{x_{i,h^\prime}}{b_{i,h^\prime}}
\end{equation}
Note that the dominate condition is formally equivalent to a sign test over a scalar product where the accumulator acts as the origin. FOCuS only tests for anomaly intervals represented by curves in the curve stack. In this sense, the curve stack serves the purpose of a dynamically updated schedule. The significance of a curve-interval is estimated according to Eq. \ref{eq:wilk_sign}. Evaluating the significance of a curve, there is no need to check for $x_{i,h}/b_{i,h} > 1$ since this is an invariant property of the algorithm.

\begin{listing} 
\begin{minted}
{C}
typedef struct {
    int x;
    double b;
    int t;
    double m;
} Curve;

double curve_max(Curve *c, Curve *acc) {
    int x = (acc->x - c->x);
    double b = (acc->b - c->b);
    assert(x > b);
    return x * log(x / b) - (x - b);
}

int curve_dominate(
    Curve* p, 
    Curve* q, 
    Curve* acc
) {
    int p_x = acc->x - p->x;
    int q_x = acc->x - q->x;
    double p_b = acc->b - p->b;
    double q_b = acc->b - q->b;
    if (p_x * q_b - q_x * p_b > 0)
        return +1;
    return -1;
}

Curve ZERO_CURVE =  (Curve) { 0 };
Curve TAIL_CURVE =  (Curve) {INT_MAX, 0., 0, 0.};
\end{minted}
\caption{Poisson-FOCuS's curves.}
\label{listing:curves}
\end{listing}

\paragraph{Constructor}
The constructor is responsible for initializing the algorithm. It pushes two curves on top of the empty curve stack. These curves do not have intrinsic meaning and their use is ancillary. We call the first of these elements the tail curve. The tail curve is always found at the bottom of the curve stack and signals the end of it. We will come back to the actual expression of the tail curve when discussing the FOCuS's update step. The second curve is the accumulator. All the members of the accumulator are initialized to zero and remain so every time the curve stack only contains the tail curve and the accumulator.
The constructor also initializes two variables for storing the algorithm's results: the global maximum and the time offset. Additionally, it computes quantities that can be calculated once and for all throughout the algorithm's operations.

\begin{listing}
\begin{minted}{C}
typedef struct {
    Stack *curves;
    double maximum;
    int time_offset;
    double mu_crit;
    double threshold;
} Focus;

void focus_init(
    Focus *f, 
    Stack *s,
    double threshold, 
    double mu_min
) {
    assert(mu_min >= 1.);
    assert(threshold > 0.);
    f->maximum = 0.;
    f->time_offset = 0;
    f->curves = s;
    f->threshold = threshold * threshold / 2;
    f->mu_crit = (
        mu_min == 1. ? 
            1. : (mu_min - 1.) / log(mu_min)
    );

    stack_push(s, &TAIL_CURVE);
    stack_push(s, &ZERO_CURVE);
}
\end{minted}
\caption{Poisson-FOCuS's constructor.}
\label{listing:init}
\end{listing}

\paragraph{Interface}
The interface accepts data from the user, executes the algorithm, and returns the results. 
Different interfaces can be adapted to different input data. Presently, we consider two input arrays representing the observation time series $X_t$ and the background time series $B_t$. In this case, the interface loops over the array indices $i$, at each step passing $x_i$ and $b_i$ to the updater. The loop goes on until the threshold is met or the signal ends. The minimal output is a triplet of numbers representing the trigger interval's end time, start time and significance. If the trigger condition is not met, a predefined signal is returned. 

Note that the updater should always be provided with a background count rate, $b_i$, greater than zero. The interface should be prepared to deal with the possibility of a bad background input and to manage the error accordingly.

\begin{listing}
\begin{minted}
{C}
typedef struct {
    int end;
    int start;
    double significance;
} Changepoint;

Changepoint focus_interface(..) {
    ..
    Focus focus;
    focus_init(&focus, &curves, threshold, mu_min);
    int i;
    for (i = 0; i < len; i++) {
        .. 
        focus_step(&focus, xs[i], bs[i]);
        if (focus.maximum > focus.threshold)
            break; // trigger
    }
    Changepoint c = {
        i, 
        i - focus.time_offset + 1, 
        sqrt(2 * focus.maximum)
    };
    return c;
}
\end{minted}
\caption{A sketch of an interface to Poisson-FOCuS.}
\label{listing:interface}
\end{listing}

\paragraph{Updater}
Through the update step, new curves are added to the curve stack, while stale ones are removed. First the algorithm creates a new curve using the latest count observation and background estimate. Then it evaluates if the new curves should be added to the curve stack, comparing it with the top-most element. 
If the new curve is not dominated by the top-most curve, it is added to the stack. Else, the algorithm discards the new curve and iterates over the stack, popping one curve at a time until a curve is found that dominates the top-most element still in the stack.
This operation is called pruning.
Finally, the last curve is evaluated and, if its mean exceeds $1$, the maximizer is invoked and the curve is pushed back on top of the stack. Otherwise the curve stack is emptied and the accumulator is reset to zero.

The accumulator is removed from the top of the stack at the very beginning of the update step, it is updated and its value is used through the function's call lifetime. The accumulator is pushed back on top of the stack just before the update step returns.
The form of the tail curve is chosen so that the pruning loop always stop when the tail curve is the top-most element in the curve stack.
Using the accumulator, Eq. \ref{eq:dominates} can be expressed:
\begin{equation}
    \frac{x_{i,i} - x_{i-h,i-h}}{b_{i,i} - b_{i-h,i-h}} > \frac{x_{i,i} - a}{b_{i,i} - b}
\end{equation}
This inequality is false when $a \rightarrow +\infty$ and $b = 0$. We can use this relation to find the right expression to the tail curve. In practice, depending on curve's representation choice, the tail curve's count member is defined to be either a  positive infinity---in the sense of floating point representation for infinities, which is supported by most modern programming languages \citep{kahan1996ieee}--or the maximum integer value, while the background member is set to zero.

According this recipe, the Poisson-FOCuS's pruning loop is equal to the hull finding step of a fundamental algorithm of computational geometry, the Graham's scan \citep{graham1972efficient}; see the classic implementations by Robert Sedgewick in Java or C \citep{sedgewick2017grahamscan, sedgewick1990algorithms}. The Graham's scan is an algorithm to efficiently determine the convex hull of a set of coplanar points. The relationship between FOCuS and convex hull algorithms has been discussed in literature before. In \cite{romano2023fast}, the FOCuS algorithm pruning step has been compared to the hull-finding step of another convex hull technique, the Melkman's algorithm.
Despite involving a double loop, the computational costs of FOCuS's update step scales linearly with the number of observations, $O(t) \propto t$. The reason is essentially the same behind the linear cost of Graham's scan hull-finding step: while the pruning loop can test curves which were first added to the curve stack much back in time, such evaluation may only occur once, since once a curve is pruned, it is never added back on the curve stack.

\begin{listing} 
\begin{minted}
{C}
void focus_step(Focus *f, int x_t, double b_t) {
    Stack *curves = f->curves;
    Curve *p = stack_pop(curves);
    Curve acc = {
        p->x + x_t, 
        p->b + b_t, 
        p->t + 1, 
        p->m
    };
    while (
        curve_dominate(p, stack_peek(curves), &acc) < 0
    )
        p = stack_pop(curves); // prune

    if ((acc.x - p->x) > f->mu_crit * (acc.b - p->b)) {
        double m = curve_max(p, &acc);
        acc.m = p->m + m;
        focus_maximize(f, p, &acc);
        stack_push(curves, p);
        stack_push(curves, &acc);
    } else {
        stack_reset(curves);
        stack_push(curves, &TAIL_CURVE);
        stack_push(curves, &NULL_CURVE);
    }
}
\end{minted}
\caption{Poisson-FOCuS's update step, with $\mu_{min}$ cut and adaptive maxima check.}
\label{listing:update}
\end{listing}

\paragraph{Maximizer}
Once the curve stack has been updated, FOCuS still needs to evaluate the significance of the intervals associated with each curve. Since the average number of curves in the curve stack grows, on average, as the logarithm of the number of iterations \citep{ward2023constant}, maximizing each curve at each iteration would result in linearithmic amortized costs, $O(t) \propto t\log(t)$. A technique called ``adaptive maxima check" was introduced in Ward et al. 2023 to speed up the maximization step \citep{ward2023constant}. This technique has been empirically observed to reduce the amortized, per-step cost of curve maximization to $O(1)$. Through the incorporation of adaptive maxima check, the amortized cost of running Poisson-FOCuS is reduced from linearithmic to linear $O(t) \propto t$, making the algorithm computationally optimal (up to a constant).

The difference between the members of two curves does not change with an update. This fact implies that at any given iteration, the global significance maximum is upper-bound by the cumulative sum of the differences between the curves in the stack. By keeping track of this upper-bound, most significance computations can be avoided. However, implementing the adaptive maxima check comes with a memory cost, as it requires adding an extra floating-point field member to the curve's definition for storing the maximum bound.

Implementing this technique is straightforward when utilizing an accumulator. Whenever a new curve is pushed on top of the stack, its maximum significance is evaluated and the accumulator's maximum bound field is incremented by that value. Then the significance of curves deeper in the curve stack is evaluated, if the trigger condition is not met and the curve's maximum bound exceeds the threshold. The iterations stop when a curve with maximum bound lower than threshold is found or the trigger condition is met. Most often, particularly for large threshold values, the loop stops at the first iteration.

\begin{listing} 
\begin{minted}
{C}
void focus_maximize(Focus *f, Curve *p, Curve *acc) {
    Stack *curves = f->curves;
    double m = acc->m - p->m;
    int i = curves->head;
    while (m + p->m >= f->threshold) {
        if (m >= f->threshold) {
            f->maximum = m;
            f->time_offset = acc->t - p->t;
            break;
        }
        i == 0 ? i = curves->capacity : i--;
        p = (curves->arr + i);
        m = curve_max(p, acc);
    }
}
\end{minted}
\caption{Poisson-FOCuS's maximizer through adaptive maxima check.}
\label{listing:maximizer}
\end{listing}

\paragraph{Memory requirements and $\mu_{\text{min}}$ cut}
\label{sec:mumin}
The memory required by FOCuS grows with the number of elements in the curve stack, hence on average as $\log(t)$ after $t$ iterations. To limit memory usage, the most obvious solution is to remove the oldest curve in the stack when the curve number exceeds a pre-fixed number. A more controllable approach requires to inhibit triggers from anomalies with mean smaller than a user-defined value $\mu_{\text{min}}$, with $\mu_{\text{min}} > 1$. In practical implementation, this requires modifying the update step method to only push a new curve on the stack if its mean exceeds the critical value:
\begin{equation}
     \mu_{\text{crit}} = \frac{\mu_{\text{min}} - 1}{\log(\mu_{\text{min}})}
\end{equation}
The critical mean value can be computed once and for all, at the constructor's level.

The algorithm's sensitivity to long, faint transients is reduced when implementing $\mu_{\text{min}}$ cut. This is often desirable in real application, due to features of the searched transients, or limits of the background estimate, or other factors. In these cases, implementing  $\mu_{\text{min}}$ cut may result in less false positives, without losing any real events. For many applications, this is more important than the memory benefit, since the memory usage of Poisson-FOCuS is very low regardless. As with the threshold parameter, there is no one-size-fits-all recipe to select the right $\mu_{\text{min}}$ value, as this value will depend on the specific application, and experimentation with the data is recommended.

\section{Tests with synthetic data}
\label{sec:synth_test}
In this section, the performances of different trigger algorithms are compared over synthetic data, targeting different performance metrics such as true positive rates and computational efficiency. The main reason for testing algorithms over simulated data is that everything about the simulation is known to the tester. For example, the true background mean rate and the actual background distribution are known, as well as the transient's timing and intensity. Using this information, it is possible to define a benchmark with ideal detection performances, setting a reference against which the performances of real techniques are evaluated. 

\begin{figure}
    \includegraphics[width=\linewidth]{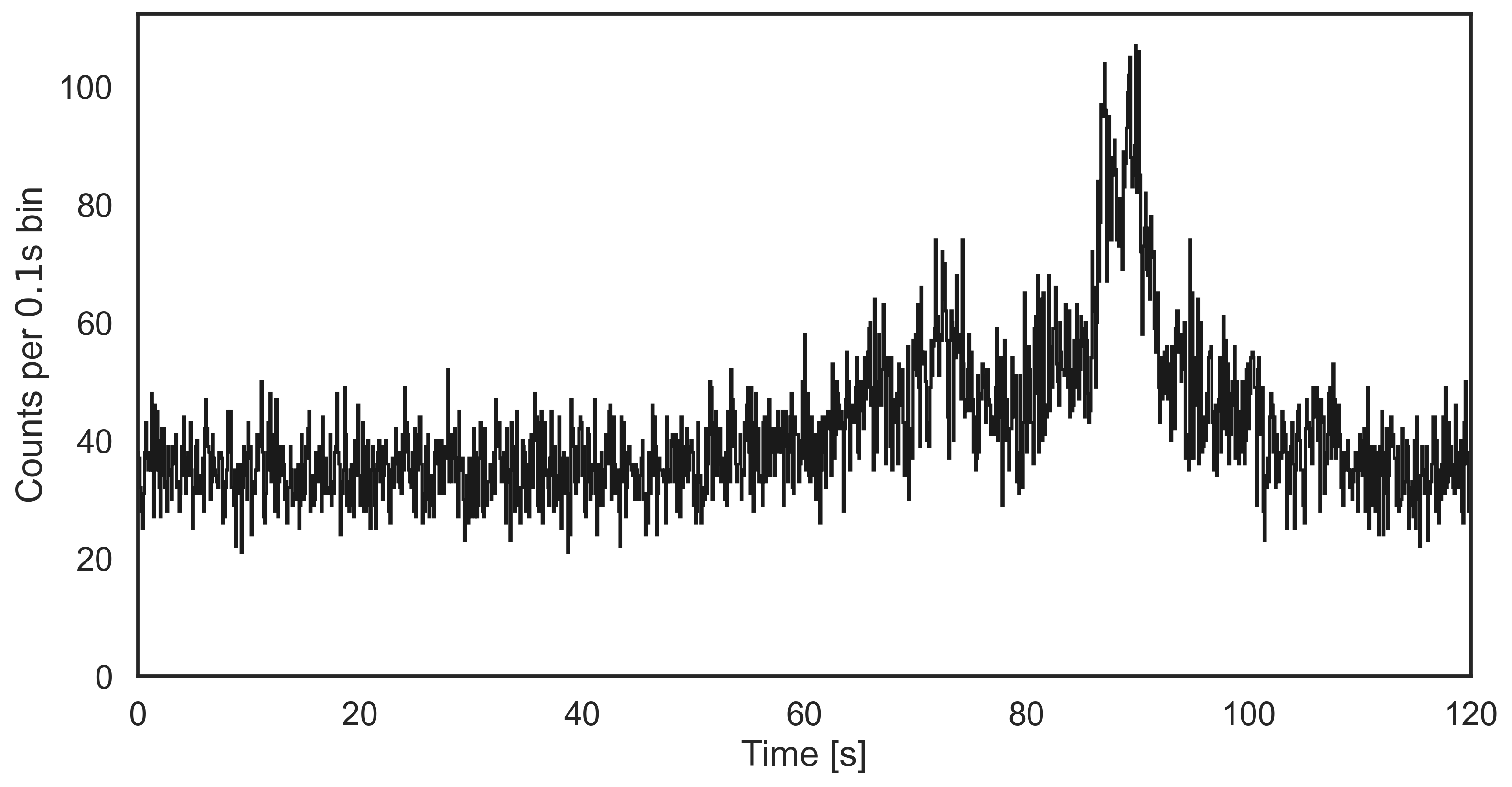}
    \caption{
    A synthetic lightcurve generated with Synthburst, using GRB120707800 as a template. Simulated data comprise $5000$ source photons and a background count rate equal $350$ photons per second.
    }
    \label{fig:sim_burst}
    
    \includegraphics[width=\linewidth]{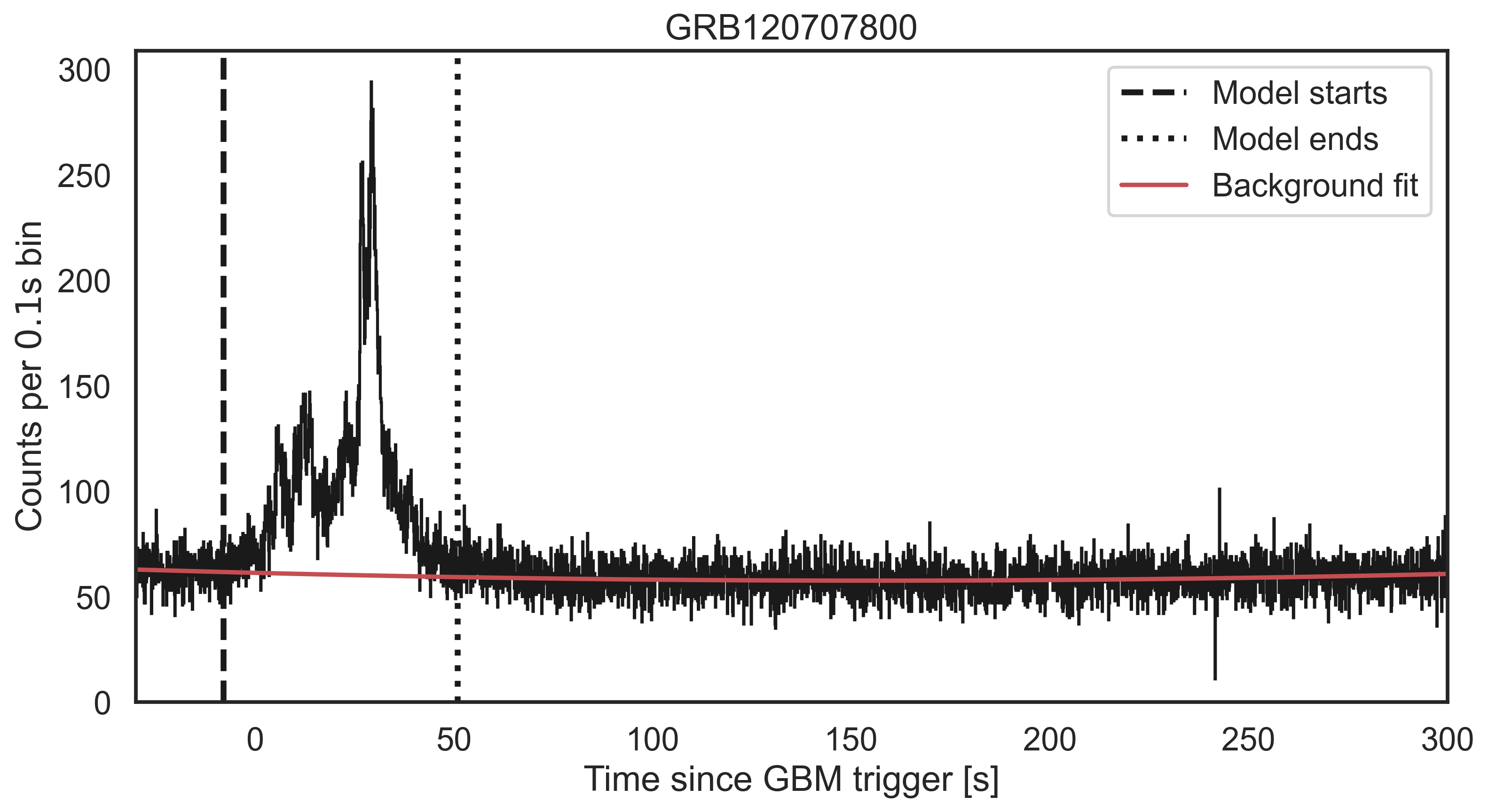}
    \caption{
    Histograms of Fermi-GBM photon counts for GRB120707800, in band $50$-$300$ keV. 
    Data comprise observations from NaI detectors $8$ and $10$. 
    Time is expressed from Fermi-GBM's trigger time. 
    Vertical lines represent the start and end time of the interval used as a template for the simulations. The background estimate (solid red line) was obtained fitting a polynomial to the observations outside this section.
    }
    \label{fig:real_burst}
\end{figure}

\subsection{Detection performances}
\label{sec:detrates}

The detection performances of different algorithms were tested and compared over two synthetic datasets. Each dataset is composed of count data from $30000$ lightcurves binned at $16$~ms, all with equal duration and mean background count rate. The background count rate is constant and equal to $350$~$s^{-1}$, a value representative of the background flux observed by the Fermi-GBM's NaI detectors in the energy band $50$-$300$ keV. Each lightcurve hosts a simulated GRB event with intensity spanning $30$ different levels. For each intensity level, a total of $1000$ lightcurves are generated. The lightcurve of the GRBs in the first dataset are modeled after the Fermi-GBM observations of the short burst GRB180703949, while the template for the lightcurves in the second dataset is the slow-raising, long burst GRB120707800 (cf. Fig.~\ref{fig:real_burst}).

To generate these data a software called Synthburst was specifically developed. Synthburst is a Monte Carlo tool for creating lightcurves resembling those produced in GRB observation. GRB sources are modeled after Fermi-GBM observations and every burst of the GBM burst catalog can be used as a template. The user can customize a synthetic lightcurve through a number of parameters. These include the source's intensity and timing, and the background's mean count rate and time profile. In Fig.~\ref{fig:sim_burst} we show a simulated lightcurve obtained using the long burst GRB120707800 as a template, see  Fig.~\ref{fig:real_burst}. Synthburst will be presented in a future publication.

We evaluated the detection performance of five distinct algorithms.  Two of these algorithms, an exhaustive search algorithm and Poisson-FOCuS, were given information on the simulation's true background count rate. The exhaustive search algorithm does not approximate the interval's significance score, rather, it computes the Poisson's tail probability and converts it to standard deviations exactly. It serves as a benchmark for detection power. In real applications, the true background value is unknown and must be estimated, often from the same data which are tested for an anomaly. The other algorithms autonomously assess the background count rate, see Appendix \ref{apx:background}. One of these is an implementation of Poisson-FOCuS with automatic background assessment through single exponential smoothing (FOCuS-AES). The remaining two emulate the algorithms of Fermi-GBM \citep{paciesas2012fermi} and BATSE \citep{1999paciesas, 1999kommers}. Details on these algorithms are provided in Appendix \ref{apx:algodetails}.

\begin{figure}
    \centering
    \includegraphics[width=\linewidth]{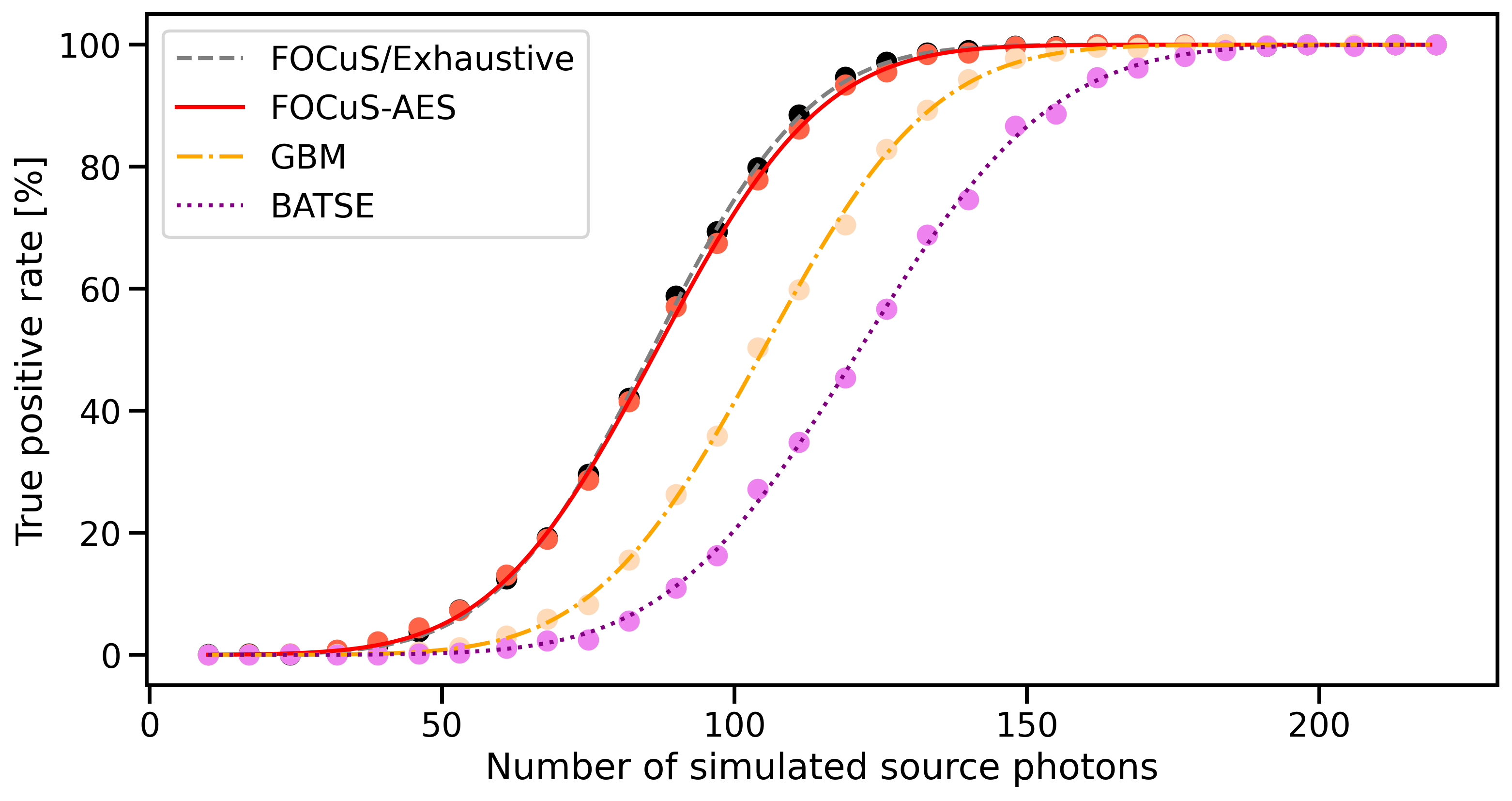}
    \caption{
    True positive rates as a function of source intensity (points) and  best fit to error function (lines) over the short GRB dataset, see the discussions of Sec.~\ref{sec:detrates} and Sec.~\ref{sec:discussion}.\\
    }
    \label{fig:deteff_short}

    \includegraphics[width=\linewidth]{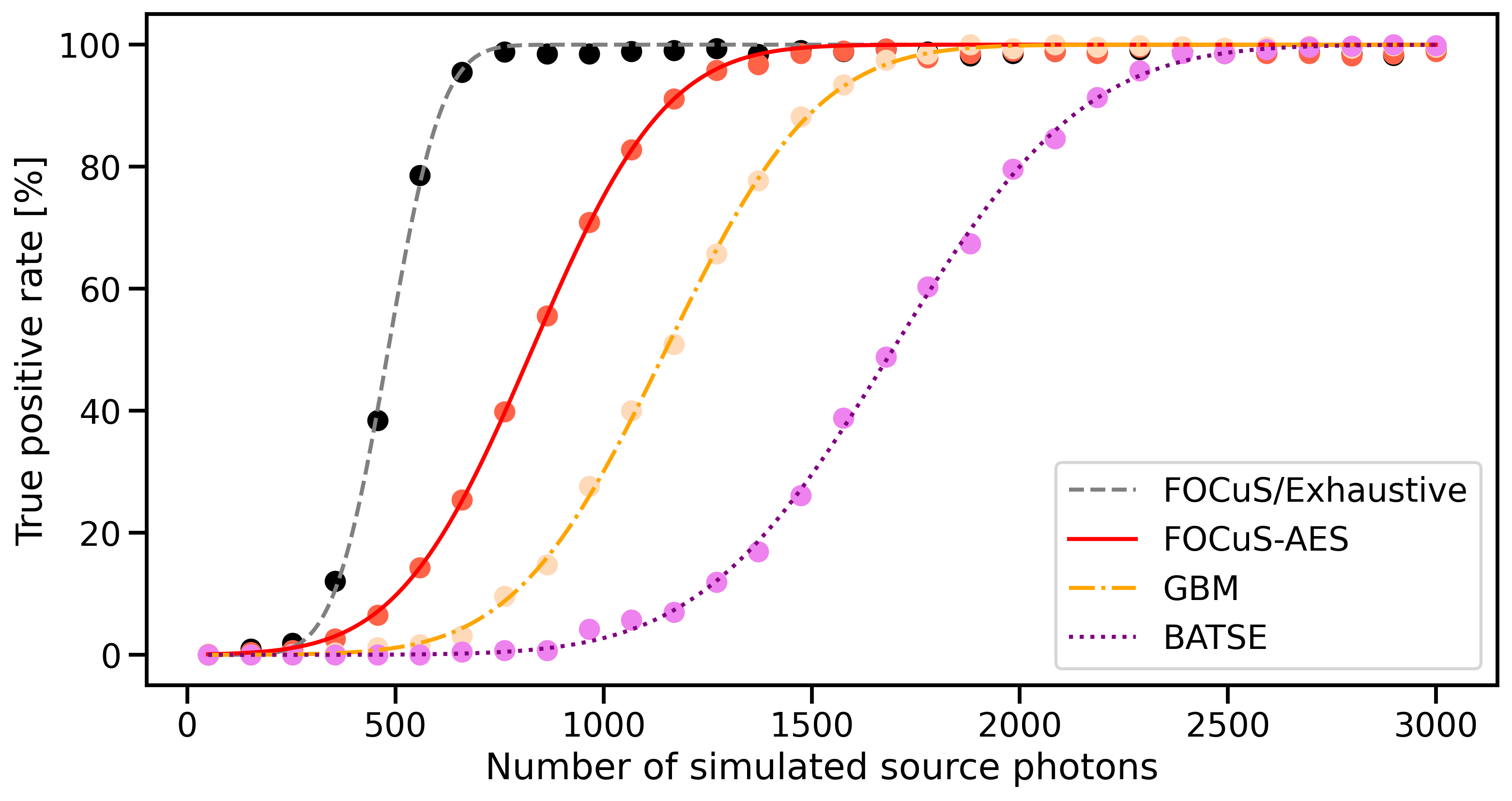}
    \caption{
    True positive rates as a function of source intensity (points) and  best fit to error function (lines) over the long GRB dataset, see the discussions of Sec.~\ref{sec:detrates} and Sec.~\ref{sec:discussion}.\\
    }
    \label{fig:deteff_long}
\end{figure}

Each trigger algorithm is executed over each lightcurve in the datasets, with threshold set to $5 \sigma$.
Lightcurve simulations and trigger execution are performed in two steps. First, the background photons are simulated. The algorithms are launched a first time and, if a trigger occurs, the event is recorded and marked as a false positive. If no false positive is observed, source photons are generated on top of the background and the algorithms are executed a second time. If an algorithm triggers, the event is recorded and marked as a true positive. Else, it is marked as false negative.
The observed rate of true positives over the short and long GRB datasets are displayed as a function of the simulated photons number in Fig.~\ref{fig:deteff_short} and Fig.~\ref{fig:deteff_long}, respectively. A summary on the relative and absolute performance of different algorithms is provided in Tab.~\ref{tab:short_tab} and Tab.~\ref{tab:long_tab}.

\begin{table*}
\centering
\begin{tabular}{lrrrrrrrr}
\toprule
 & & & & \multicolumn{5}{c}{Relative efficiency [\%]} \\ \cmidrule(r){5-9}
& $t_p$ & $f_p$ & $f_n$ & $F_0$ & $F_1$ & $F_2$ & $F_3$ & $F_4$ \\
\midrule
Exhaustive & 19024 & 26 & 10950 & 50.0 & 52.0 & 51.5 & 81.5 & 95.3 \\
FOCuS & 18877 & 16 & 11107 & 48.0 & 50.0 & 49.5 & 80.2 & 94.8 \\
FOCuS-AES & 18913 & 23 & 11064 & 48.6 & 50.5 & 50.0 & 79.4 & 94.0 \\
GBM & 16400 & 9 & 13591 & 20.3 & 21.6 & 21.3 & 50.0 & 76.4 \\
BATSE & 14100 & 4 & 15896 & 8.5 & 9.2 & 9.0 & 26.3 & 50.0 \\
\bottomrule
\end{tabular}
\caption{Summary of detection efficiency tests over the short GRB model dataset, see Sec.~\ref{sec:real_test} and Appendix \ref{apx:algodetails}. Results from different algorithms are presented in separate rows. The first three columns indicate the total number of true positives ($t_p$), false positives ($f_p$), and false negatives ($f_n$). In the last five columns, the true positive rates are expressed as percentages for five different source intensity levels ($F_i$). Specifically, $F_0$ represents the source level at which \emph{Exhaustive} exhibited a 50\% false positive rate, while $F_1$, $F_2$, $F_3$ and $F_4$ correspond to the intensity levels at which \emph{FOCuS}, \emph{FOCuS-AES}, \emph{GBM}, and \emph{BATSE} achieved the same sensitivity level, respectively.} 
\label{tab:short_tab}
\end{table*}

\begin{table*}
\centering
\begin{tabular}{lrrrrrrrr}
\toprule
 & & & & \multicolumn{5}{c}{Relative efficiency [\%]} \\ \cmidrule(r){5-9}
& $t_p$ & $f_p$ & $f_n$ & $F_0$ & $F_1$ & $F_2$ & $F_3$ & $F_4$ \\
\midrule
Exhaustive & 25001 & 351 & 4648 & 50.0 & 50.8 & 100.0 & 100.0 & 100.0 \\
FOCuS & 25072 & 230 & 4698 & 49.2 & 50.0 & 100.0 & 100.0 & 100.0 \\
FOCuS-AES & 21626 & 368 & 8006 & 8.6 & 8.7 & 50.0 & 89.8 & 100.0 \\
GBM & 18668 & 89 & 11243 & 1.0 & 1.0 & 13.1 & 50.0 & 97.1 \\
BATSE & 13360 & 18 & 16622 & 0.0 & 0.0 & 0.8 & 6.6 & 50.0 \\
\bottomrule
\end{tabular}
\caption{Summary of detection efficiency tests for different algorithms over the long GRB model dataset, see Tab.~\ref{tab:short_tab}.}
\label{tab:long_tab}
\end{table*}

\subsection{Computational efficiency}
\label{sec:compeff}

We compared the computing times of Poisson-FOCuS against a similarly implemented benchmark emulator for the Fermi-GBM on-board trigger logic. The tests were performed over Poisson-distributed count time series with constant mean rate. Multiple count time series with different length and mean rate have been generated and evaluated. The algorithms we test do not assess background count rates automatically; they are informed on the true count mean rate. The rationale behind this choice is that different algorithms for anomaly detection may employ the same technique to assess the background count rate. For both algorithms, the interval significance is defined according to Eq. \ref{eq:wilk_sign}. Programs are executed without checking for the trigger condition to guarantee a complete run over the input dataset.

To our knowledge, the actual implementation of the algorithm used by Fermi-GBM has not been disclosed, nor has the code from any other major experiment for GRB detection. Hence, we developed our own implementation for benchmarking purposes, focusing on optimizing performance. We implemented three key optimizations. First, the program stores photon counts cumulatively over a circular array. This allows computing the total number of photon counts $x_{i,h} = \sum_{t=i-h+1}^{i} x_t$ through a single difference operation $I(i) - I(h)$. This optimization has been described in literature before, see \cite{fenimore2003trigger}. Second, a phase counter $p$ is employed to determine \emph{for which criteria it is the right time to perform a test}, avoiding looping over individual test criteria, at each iteration. Since the benchmark's timescales are logarithmically equispaced, one can loop over the timescales, doubling the timescale parameter $h$ until the modulo operator of $h$ and $p$ is found equal to zero or the maximum timescale is reached. 
Finally, the program performs a significance computation only if an interval's observed photon count exceeds the expected count from the background.
The benchmark's test criteria timescales and offsets are the same described in Sec.~\ref{sec:detrates} for the algorithm labeled ``\emph{GBM}". This parameter choice was intended to emulate the operations of the Fermi-GBM algorithm over a single detector in band $50-300$ keV, as described in \cite{paciesas2012fermi}.

The Poisson-FOCuS's curve stack was implemented over a circular array of fixed length and can host up to $64$ different curves. This choice was meant to avoid dynamic memory allocation. If a curve is pushed onto a full stack, the oldest curve in the curve stack is removed. The chance of this to happen is minimized by the use of a $\mu_{min}$ parameter, which was set to $1.1$. Since the Poisson-FOCuS maximization step requires a threshold value to be defined, threshold level was set to $5.0$ $\sigma$---even if, as we already stated, the halting condition is never tested during program execution.

The implementations we test are written in C99 and compiled with GCC. Tests were performed with a 3.2 GHz Apple M1 ARM processor. 
Results are reported in Tab.~\ref{tab:comp_eff}, averaging over $20$ different input dataset per entry. Across the table, the maximum observed standard deviation amounts to less than $10\%$ of the mean. Values do not account for the time required to read the input files.

\begin{table}
\centering
\begin{tabular}{lccccccccc}
\toprule
    & \multicolumn{3}{c}{Poisson-FOCuS [ms]}  & \multicolumn{3}{c}{Benchmark [ms]} \\ \cmidrule(r){2-4} \cmidrule{5-7}
$N$ & 4.0 & 16.0 & 64.0 & 4.0 & 16.0 & 64.0 \\
\midrule
2048 & 0.04 & 0.04 & 0.04 & 0.08 & 0.08 & 0.08 \\
4096 & 0.09 & 0.08 & 0.08 & 0.16 & 0.16 & 0.17 \\
8192 & 0.17 & 0.16 & 0.15 & 0.32 & 0.33 & 0.34 \\
16384 & 0.33 & 0.32 & 0.30 & 0.63 & 0.63 & 0.66 \\
32768 & 0.67 & 0.63 & 0.59 & 1.24 & 1.25 & 1.31 \\
65536 & 1.33 & 1.28 & 1.19 & 2.45 & 2.50 & 2.57 \\
131072 & 2.65 & 2.51 & 2.36 & 4.86 & 4.92 & 5.09 \\
262144 & 5.29 & 5.06 & 4.69 & 9.66 & 9.84 & 10.12 \\
524288 & 10.59 & 10.16 & 9.45 & 19.31 & 19.71 & 20.33 \\
1048576 & 21.21 & 20.24 & 18.84 & 38.65 & 39.19 & 40.48 \\
\bottomrule
\end{tabular}
\caption{Time, in milliseconds, required by Poisson-FOCuS and a benchmark emulator for the Fermi-GBM algorithm to run over  Poisson-distributed count time series with four different means ($4.0$, $16.0$, $64.0$) and geometrically increasing lengths ($N$), using a retail $3.2$ GHz ARM processor. Each reported value is the averaged over twenty randomly generated time series. For all entries the observed standard deviation is smaller than $10\%$.}
\label{tab:comp_eff}
\end{table}

\subsection{Discussion}
\label{sec:discussion}
We test whether the detection performance of FOCuS-AES are equal to that of the exhaustive search benchmark. According to the results of Tab.~\ref{tab:comp_eff}, there is not significant evidence for a difference for the short GRB dataset (p-value~$= 0.34$), but there is strong evidence for a difference in the long GRB dataset (p-value~$< 10^{-5}$). On the other hand, the detection of the Poisson-FOCuS implementation with access to the true background mean are statistically similar to that of the exhaustive search over both the short (p-value~$= 0.21$) and long (p-value~$= 0.78$) GRB datasets---as one would expect from the theory. We conclude that the detection performance of FOCuS-AES are hindered by the background estimation process over the long GRB dataset: for long, faint bursts, the background estimate may be ``polluted" by the source's photons. Nevertheless, FOCuS-AES achieved the best correct detection rates among algorithms with no access to the background's true value. We remark that Poisson-FOCuS and the benchmark resulted in a higher false positive rate when compared to alternatives. This discrepancy is due to the number of intervals effectively tested by the algorithms: FOCuS and the exhaustive search benchmark effectively test all the intervals of the time series, while conventional algorithms test a subset whose extension depends on the choice of the timescales and offsets parameters. 

In Sec.~\ref{sec:compeff}, we compared the computational performances of Poisson-FOCuS to that of a similarly implemented benchmark emulating the Fermi-GBM on-board algorithm. The time required to run Poisson-FOCuS over count series of length spanning $2048$ and $1048576$ are observed to increase linearly, as evident by comparing column-wise values in Tab.~\ref{tab:comp_eff}, regardless of the mean rate. Our implementation required less than $22$ ms to analyze over one million data using a retail $3.2$ GHz ARM processor. For comparison, the benchmark algorithm required approximately double the time in all the tests we performed. 
We remark that our benchmark is not the routine actually implemented on-board Fermi-GBM, as this implementation remains undisclosed. Hence, the reader should avoid interpreting the results of this test as a direct comparison with the algorithm of Fermi-GBM.

\section{Tests with real data}
\label{sec:real_test}
We ran Poisson-FOCuS with automatic background assessment by simple exponential smoothing (FOCuS-AES) over three weeks of data from Fermi-GBM, chosen at different times during the mission.
The algorithm's implementation and parameters are the same specified in Sec.~\ref{sec:detrates}. We implemented a halting condition similar to the one of Fermi-GBM: a trigger is issued whenever at least two detectors overcome a $5\sigma$ threshold; after a trigger, the algorithm is kept idle for $300$~s after which it is restarted.  
Input data were obtained binning the Fermi-GBM daily TTE data products \citep{fermidataproduct} at $16$ ms time intervals over the energy band  $50-300$ keV. Only data from the GBM's $12$ NaI detectors were analyzed. This means that data from the GBM's bismuth germanate (BGO) detectors and photons with low ($< 50$ keV) and high ($> 300$ keV) energy were not tested. The reason for this choice is twofold. First, these data pertains to regimes marginally relevant to GRB detection--see for example Tab.~4 and relative discussion from \cite{von2014second}. Second, the threshold profile employed by the Fermi-GBM algorithm outside the $50$ - $300$ keV domain is highly fragmented---see Tab.~2 of \cite{von2014second} or \cite{von2020fourth}---which complicates the setup of a meaningful comparison.

The first period we tested spans the week from 00:00:00 2017-10-02 to 23:59:59 2017-10-09 UTC time. 
There were two factors taken into account when choosing this time period. One factor was its association with multiple, reliable short GRB candidates, as reported in the untriggered GBM Short GRB candidates catalog \citep{fermiuntriggered}, which had gone undetected by the Fermi-GBM's online trigger algorithm. The other factor was its coincidence with a period of moderate solar activity.
For this week the Fermi-GBM's Trigger Catalog \citep{fermidataproduct} reports eleven detections. Six of these events were classified as GRBs, three as terrestrial gamma-ray flashes (TGF), one as a local particle event and one as an ``uncertain" event.
FOCuS-AES triggered eight times, correctly identifying all the GRBs and local particle events detected by the Fermi-GBM algorithm. Another event triggered FOCuS-AES at time 16:05:52 2017-10-02 UTC. This event is consistent with one of the transients listed in the untriggered GBM Short GRB candidates catalog at Fermi's MET time 528653157.
No counterpart to the uncertain event UNCERT17100325 was identified. The three TGF detected by the Fermi-GBM's on-board algorithm went also undetected. This is understandable: TGF are more energetic than GRBs and are detected either over the BGO detectors---which are sensitive over an energy range spanning from approximately $200$~keV to $40$~MeV---or the NaI detector at energy range $> 300$~keV. The former was the case for the events of this sample, TGF171004782, TGF171008504 and TGF171008836.

The second period spans the week from 00:00:00 2019-06-01 to 23:59:59 2019-06-08 UTC time, a period of low solar activity in which the Fermi-GBM on-board algorithm triggered nine times. Seven of these events were classified as GRBs, one as a local particle event and one as a TGF. FOCuS-AES detected ten events. All the GRBs, as well as the local particle events detected by Fermi-GBM, were identified. No counterpart was found to TGF190604577--again, this is understandable, since we did not test the data over which TGFs are detected. The remaining two events detected with FOCuS-AES are compatible with events in the untriggered GBM Short GRB candidates catalog at MET time 581055891 and 581117737.

The last week occurred during a period of high solar activity, from 00:00:00 2014-01-01 to 23:59:59 2014-01-08 UTC time. For this week, the Fermi-GBM Trigger Catalog reports 29 detections: six GRBs, two local particle events, one TGF, one uncertain event and nineteen solar flares. In our test, FOCuS-AES got 36 positive detections. All GRBs discovered by Fermi-GBM were correctly identified. We did not find counterparts for six solar flares, two local particles, one TGF and one uncertain event. All these events were either low-energy or detected over Fermi-GBM's BGO detectors. FOCuS-AES detected 16 events with no counterparts in the Fermi-GBM Trigger Catalog nor the untriggered GBM Short GRB candidates catalog, see Tab.~\ref{tab:unknownevents} of the appendix. It is not immediately apparent that any of these occurrences are false positives. Indeed, each of these events corresponds to short transients that are readily identifiable, with high temporal coincidence, over multiple detectors and different energy ranges. However, due to the frequency and clustering of these events we expect most of them to be of solar origin.

\section{Conclusions}
In this paper:
\begin{enumerate}
    \item We discussed a framework to evaluate and visualize the operations of different algorithms for detecting GRBs.
    \item We presented an implementation of Poisson-FOCuS that was designed and proposed for application on the satellites of the HERMES-Pathfinder constellation.
    \item We tested the detection performances of different algorithms using simulated data. We found that realistic implementations of Poisson-FOCuS have higher detection power than conventional strategies, reaching the performance of an ideal, exhaustive search benchmark when not hindered by automatic background assessment.
    \item We tested the computational efficiency of Poisson-FOCuS against that of a similarly implemented benchmark emulator for the Fermi-GBM on-board trigger logic. In these tests, Poisson-FOCuS proved twice faster than the benchmark.
    \item We validated Poisson-FOCuS with automatic background assessment over three weeks of data from Fermi-GBM.
\end{enumerate}
These findings highlight the potential of Poisson-FOCuS as a tool for detecting GRBs and other astrophysical transients, especially in resource-constrained environments such as nanosatellites. However, the effectiveness of a trigger algorithm is limited by the accuracy of background estimates. In the future, machine learning may provide a reliable solution to this challenge. Poisson-FOCuS achieves high detection power by effectively testing all intervals within a count time series for anomalies. This approach also leads to a higher rate of false positives. When selecting an algorithm, it is important to consider this trade-off. Clever instrument design can mitigate this issue, providing ways to assure the quality of a trigger.
While Poisson-FOCuS was designed for the time domain, offline changepoint models for detecting variability over both times and energies exist in the literature \citep{wong2016detecting}. It is possible that a FOCuS-like algorithm could be devised for the multidimensional setting, making similar techniques suitable also for online applications. Such an effort is beyond the scope of this work, but presents a promising direction for future research.

\clearpage
\section{Code availability statement}
The code used in this research has been made public at \url{https://github.com/peppedilillo/grb-trigger-algorithms} and \url{https://doi.org/10.5281/zenodo.10069414} \citep{giuseppe_dilillo_2023_10069414}. All the data used in this research are available at \url{https://doi.org/10.5281/zenodo.10034655} \citep{giuseppe_dilillo_2023_8334676}. 

\begin{acknowledgments}
We acknowledge support from the European Union Horizon 2020 Research, and Innovation Framework Programme under grant agreement HERMES-Scientific Pathfinder n. 821896; and Horizon 2020 INFRAIA Programme under Grant Agreement n. 871158 AHEAD2020; and from ASI-INAF Accordo Attuativo HERMES Technologic Pathfinder n. 2018-10-H.1-2020; and INAF RSN-5 mini-grant 1.05.12.04.05, ``Development of novel algorithms for detecting high-energy transient events in astronomical time series"; and EPSRC grant EP/N031938/1.
This research has made use of data provided by the High Energy Astrophysics Science Archive Research Center (HEASARC), which is a service of the Astrophysics Science Division at NASA/GSFC. We thank the anonymous reviewers for their precious insights and feedback. Giuseppe Dilillo thanks Riccardo Campana for his mentoring.

\software{
    Astropy \citep{astropy:2013},
    Numpy \citep{harris2020array},
    Pandas \citep{reback2020pandas},
    Matplotlib \citep{Hunter:2007},
    Scipy \citep{2020SciPy-NMeth},
    LMFIT \citep{newville_matthew_2014_11813},
    Joblib \citep{joblib}
}
\end{acknowledgments}

\appendix

\section{Background Estimate}
\label{apx:background}

In this section, we discuss the problem of assessing an estimate of the background count rate. With count detectors on-board spacecrafts, this task is complicated by two factors: 1. the fact that the background mean count rate changes over time, although with characteristic timescales generally larger than the duration of most GRBs; and 2. the need to evaluate the background estimate from the same data that are tested for anomalies. 

\paragraph{Background time-dependence} The background count rate of a space-borne count detector is influenced by multiple phenomena such as the solar activity, the rate of incident primary and secondary cosmic ray particles, the X-ray and gamma-ray diffuse photon background, trapped radiation, nuclear activation and the instrument's intrinsic noise \citep{campana2013background}. The interaction between these phenomena is complex and can result in background counts that fluctuate considerably over time. In Low-Earth orbits, periodic fluctuations exist on timescales equal to the orbit duration (approximately 90 minutes at 500 km altitude) and the duration of a day due to features of the orbital geomagnetic environment and the spacecraft's motion. 
The magnitude and velocity of the fluctuations depend on the orbital inclination parameter. Spacecraft in near-equatorial orbits span a limited range of geomagnetic latitudes and only graze the outermost regions of the South Atlantic Anomaly (SAA), resulting in small and regular background modulations due to the detector's attitude and the residual variation of the geomagnetic field \citep{campana2013background}.
In contrast, the background count rate of a spacecraft crossing the SAA in depth may change significantly, even on short timescales. 

\paragraph{Moving average techniques} The most common technique employed by GRB monitor experiments to estimate the background count rate is the simple moving average. With simple moving average, past data are summarized by the unweighted mean of $n$ past data points:
\begin{equation}
M_{t,n} = \frac{x_{t,n}}{n}
\end{equation}
Which may be efficiently computed through the recursion:
\begin{equation}
\label{eq:sma_fast}
b_{t} = b_{t} + \frac{x_{t} - x_{t-n+1}}{n}
\end{equation}
Simple moving averages are suitable for assessing background estimates if $n$ is larger than the GRB characteristic rise time and the background count rates change over timescales larger than $n$. The evaluation can be delayed to reduce `pollution' from eventual source photons, hence a higher rate of false negatives. For example, this is the approach of Fermi-GBM, which estimates the background count rate over a period nominally set to $17$ s, excluding the most recent $4$ s of observations from the computation \citep{meegan2009fermi}. Computing a simple moving average requires maintaining one (two, if the evaluation is delayed) queue buffers of recent counts observations.
Exponential smoothing is an alternative to moving average techniques, see \cite{hyndman2018forecasting} for an overview. In single exponential smoothing, exponentially decreasing weights are assigned to past observations. At time-index $t$, the single exponential smoothing estimate is:
\begin{equation}
\label{eq:SES}
b_t = \alpha x_{t} + (1 - \alpha) b_{t-1}
\end{equation}
where $0 < \alpha \leq 1$ is the smoothing constant parameter.  The choice of the $\alpha$ parameter depends on the features of the data and is generally achieved optimizing for the mean squared error over a grid of parameters over a test dataset. When $\alpha$ is close to $0$ more weight is given to older observations, resulting in slower dampening and smoother $S_t$ curves. On the other hand, $\alpha$ value near unity will result in quicker dampening and more jagged $S_t$ curves.  Computing $S_t$ requires the setting of an initial value $S_0$. Common initialization techniques include setting $S_0$ to $x_0$, to an a priori estimate of the process target, or to the average of a number of initial observations. Similarly to simple moving averages, exponential smoothing estimates can be delayed to prevent contamination from source photons. Simple exponential smoothing requires no (one with delayed evaluation) count queue buffer to be maintained.
Single exponential smoothing techniques can be modified to include a second constant accounting for trends in the data. The resulting technique is called double exponential smoothing and requires computing the recursions:
\begin{equation}
\label{eq:DES}
\begin{split}
s_t &= \alpha x_t + (1 - \alpha)(s_{t-1} + d_{t-1})\\
d_t &= \beta (s_t - s_{t-1}) + (1 - \beta) d_{t-1}
\end{split}
\end{equation}
where $0 < \alpha \leq 1$ and $0 < \beta \leq 1$. The first smoothing equation adjusts $s_t$ for the weighted trend estimate observed during the last iteration $d_{t-1}$. Common initialization techniques for the trend parameter include setting $b_0$ to $x_2 - x_1$ or an average of the differences between initial subsequent pairs of observations. At step index $t$, the $m$-step ahead forecast is given by:
\begin{equation}
\label{eq:des_forecast}
b_{t} = s_{t - m} + m d_{t - m}
\end{equation}
Since trends are indeed present in background data from count rate detectors, the application of double exponential smoothing to the problem of detecting GRBs looks promising. However, we were unable to achieve satisfying results using this technique. We found that algorithms using double exponential smoothing were more prone to false detection and harder to optimize than counterparts based on simple moving averages and single exponential smoothing.
This instability is due to background estimates ``lagging" behind the true value. This issue is not unique to double exponential smoothing and is encountered with the other techniques discussed thus far. Yet, accounting for trends appears to further exacerbate this problem. For simple moving averages, the lagging behavior of the background estimate could be addressed by ``bracketing" the interval tested for GRB onset between two regions where the background count rate is actually estimated. This solution has been pursued with the long rate trigger algorithms of Swift-BAT \citep{fenimore2003trigger}. This approach poses at least two problems. The first is that bracketing the test interval results in a delayed detection by a time duration equal to the length of the fore bracket. The second problem is that the background estimate would become destabilized when it is most needed, i.e., when the anomaly enters the test interval, passing through the fore bracket.

Physical models with excellent performance in modeling the background have been described in the literature \citep{biltzinger2020physical}. It is uncertain whether these techniques are applicable to online search, especially under constraints of limited computational resources. Machine learning models have been proven effective in modeling background count rates from space-borne detectors \citep{crupi2023searching} and can be implemented on devices with modest computational capability after training \citep{david2021tensorflow}. However, such models still require training on large datasets, a resource that is not available before the deployment of a mission.

\section{Algorithm details}
\label{apx:algodetails}
In this section we give details on the implementation of the algorithms tested in Sec.~\ref{sec:detrates}.

\paragraph{Exhaustive search}
An exhaustive search algorithm (labeled  \emph{Exhaustive}), see Sec.~\ref{sec:exh}. This algorithm was given access to true background count rate and computes significance scores exactly. Unconcerned with computational efficiency, it was designed to provide a standard reference for the detection power.

\paragraph{Poisson-FOCuS}
An implementation of Poisson-FOCuS (labeled  \emph{FOCuS}), as described to length in Sec.~\ref{sec:focuspois}, with access to true background count rate (labeled  \emph{FOCuS}) and no $\mu_{\text{min}}$ cut (i.e., $\mu_{\text{min}} = 1$). 

\paragraph{FOCuS-AES}
An implementation of Poisson-FOCuS with $\mu_{min}$ cut and automatic background assessment (labeled  \emph{FOCuS-AES}). Background mean rate is estimated through single exponential smoothing with smoothing parameter $\alpha = 0.002$~s, excluding the most recent $4.0$~s of observation. The background estimate is updated at each algorithm iteration. The exponential smoothing parameter $\alpha$ is initialized to the mean of the count observed in the first $16.992$ s of observations. Intervals with duration longer than $4.0$~s are not tested for triggers.  This precaution avoids testing data which have been used to assess background. The $\mu_{min}$ parameter is set to $1.1$.

\paragraph{GBM}
An algorithm designed to emulate the Fermi-GBM's on-board trigger over band $50$-$300$ keV (labeled  \emph{GBM}), see \cite{paciesas2012fermi}.
Background is automatically assessed through a simple moving average with length $16.992$~s, at each iteration and excluding the most recent $4.0$~s of observations. The algorithm checks nine logarithmically equispaced timescales equivalent to $0.016$~s, $0.032$~s, $0.064$~s, $0.128$~s, $0.256$~s, $0.512$~s, $1.024$~s, $2.048$~s and $4.096$~s. For all but the $0.016$~s and $0.032$~s timescales, checks are scheduled with phase offset equal to half the accumulation length e.g., a timescale with characteristic length $1.024$s, is checked $4$ times in $2.048$~s. The significance scores are defined according to Eq. \ref{eq:wilk_sign}. Two reasons drove this choice. First, to our knowledge, the exact formula implemented by the Fermi flight-software to compute significance scores has not been described in the literature. Second, this choice improves comparability with the Poisson-FOCuS. In this regard, it's worth noting that Poisson-FOCuS can be modified to compute significances according to arbitrary recipes. 

\paragraph{BATSE}
An algorithm emulating the Compton-BATSE's on-board logic (labeled  \emph{BATSE}). 
The algorithm checks three timescales equivalent to $0.064$~s,  $0.256$~s,  $1.024$~s. Differently from \emph{GBM}, only non-overlapping intervals are tested. For example, the $1.024$~s timescale is tested $2$ times in $2.048$~s. 
As with the GBM emulator, the background is automatically assessed through a simple moving average with length $16.992$~s, at each iteration and excluding the most recent $4.0$~s of observations. Significance is computed according to Eq. \ref{eq:wilk_sign}.

\section{Detections with no counterparts in Fermi-GBM catalogs}
\begin{table}[h]
\centering
\begin{tabular}{rrrr}
\toprule
\multicolumn{2}{c}{Trigger time} & & \\ \cmidrule(r){1-2}
MET & UTC & $\tau$ [ms] & Detectors \\
\midrule
410299390.55 & 2014-01-01 20:03:07 & 48	 & 6, 8 \\
410299746.18 & 2014-01-01 20:09:03 & 48	 & 6, b \\
410304166.77 & 2014-01-01 21:22:43 & 48	 & 7, 8 \\
410323479.59 & 2014-01-02 02:44:36 & 32	 & 6, 7 \\
410357487.64 & 2014-01-02 12:11:24 & 3168 & 1, 2 \\
410555917.41 & 2014-01-04 19:18:34 & 64	 & 5, 8 \\
410556407.32 & 2014-01-04 19:26:44 & 80	 & 8, a \\
410556762.28 & 2014-01-04 19:32:39 & 48	 & 7, 9 \\
410557112.25 & 2014-01-04 19:38:29 & 64	 & 7, 9 \\
410563178.79 & 2014-01-04 21:19:35 & 80	 & 7, 5 \\
410568048.55 & 2014-01-04 22:40:45 & 32	 & 2, 5 \\
410568417.21 & 2014-01-04 22:46:54 & 64	 & 8, b \\
410667708.04 & 2014-01-06 02:21:45 & 64	 & 4, 6, 7, 8 \\
410812804.74 & 2014-01-07 18:40:01 & 48	 & 1, 4, 9 \\
410813905.69 & 2014-01-07 18:58:22 & 96	 & 6, 8, 9 \\
410819366.44 & 2014-01-07 20:29:23 & 96	 & 6, 7, 8 \\
\bottomrule
\end{tabular}
\caption{Trigger events with no counterparts in the Fermi-GBM Trigger Catalog \citep{fermidataproduct} and the untriggered GBM Short GRB candidates catalog \citep{fermiuntriggered}, see the discussion of Sec.~\ref{sec:real_test}. The first two columns report the trigger times, in Fermi MET and UTC time standards. The third column holds the transient's rise timescale $\tau$ in units of milliseconds. Finally, the last column reports the detectors where the transients were detected. We note the Fermi-GBM's NaI detectors number $10$ and $11$ with the letters $a$ and $b$, respectively.}
\label{tab:unknownevents}
\end{table}

\clearpage
\section{Poisson-FOCuS minimal functional implementation}
\label{apx:minimalfocus}
\begin{listing}[h]
\begin{minted}
{python}
from math import log


def curve_update(c, x_t, b_t):
    return c[0] + x_t, c[1] + b_t, c[2] + 1

def curve_max(c):
    return c[0] * log(c[0] / c[1]) - (c[0] - c[1])

def dominates(c, k):
    return c[0] / c[1] > k[0] / k[1]

def focus_maximize(cs):
    return max(
        [(c[0] and curve_max(c) or 0, c[2]) for c in cs]
    )

def focus_update(cs, x_t, b_t, c):
    if cs and dominates(
        k := curve_update(cs[0], x_t, b_t), c
    ):
        return [k] + focus_update(cs[1:], x_t, b_t, k)
    return [(0, 0.0, 0)]

def focus(xs, bs, threshold):
    cs = [(0, 0.0, 0)]
    for t, (x_t, b_t) in enumerate(zip(xs, bs)):
        if b_t <= 0: raise ValueError()
        cs = focus_update(cs, x_t, b_t, (1, 1.0, 0))
        global_max, time_offset = focus_maximize(cs)
        if global_max > threshold:
            return (global_max, t - time_offset + 1, t)
    return 0.0, len(xs) + 1, len(xs)
\end{minted}
\caption{A minimal, functional implementation of Poisson-FOCuS in Python 3.8. Does not implement any optimization, namely accumulator, $\mu_{\text{min}}$ cut and maxima adaptive check.}
\label{listing:minimalfocus}
\end{listing}

\clearpage
\bibliography{bibliography.bib}{}
\bibliographystyle{aasjournal}

\end{document}